# Surface magnetic states of Ni nanochains modified by using different organic surfactants

Weimeng Chen[1], Wei Zhou[2], Lin He[1], Chinping Chen[1], and Lin Guo[2]

[1]Department of Physics, Peking University, Beijing, 100871, People's Republic of China

[2]School of Chemistry and Environment Science, Beijing University of Aeronautics and Astronautics, Beijing, 100191, People's Republic of China

E-mail: cpchen@pku.edu.cn and guolin@buaa.edu.cn

## Abstract

Three powder samples of Ni nanochains formed of polycrystalline Ni nanoparticles with an estimated diameter of about 30 nm have been synthesized by a wet chemical method using different organic surfactants. These samples, having magnetically/structurally core-shell structures, all with a ferromagnetic Ni core, are Ni@Ni$_3$C nanochains, Ni@Ni$^{SG}$ nanochains with a spin glass (SG) surface layer, and Ni@Ni$^{NM}$ nanochains with a nonmagnetic (NM) surface layer. The average thickness of the shell for these three samples is determined as about 2 nm. Magnetic properties tailored by the different surface magnetism are studied. In particular, suppression in saturation magnetization usually observed with magnetic nanoparticles is revealed to arise from the surface magnetic states with the present samples.

## 1. **Introduction**

Surface magnetic states have long been a subject of intense studies in the past few decades. It becomes increasingly important with the emerging field of nanomagnetics. For biomedical applications, magnetic nanoparticles play an important role for the bounding of antibiotics, nucleotides, vitamins, peptides, etc. [1, 2, 3]. These depend much on the magnetic properties of the nanoparticles, especially, the surface magnetism. In information storage, the surface anisotropy plays an important role. It offers an extra degree of freedom to tune the magnetic anisotropy energy, making it attractive for basic investigation [4, 5]. The fundamental role of exchange bias, widely counted on in spin valve and tunneling devices, is directly related to the interaction from the interface or surface [6]. Therefore, the investigations of surface magnetic states are interesting and of great importance.

Theoretical and experimental works reveal that surface atoms may have either enhanced or quenched moments, depending on their chemical environment [7, 8]. Enhancement of saturation magnetization, $M_S$, has been reported in small-sized, elemental metallic clusters [9]. However, many more thin films and nanoparticles demonstrate otherwise [10]. It is reported that the coating of organic molecules [11], CO chemisorption [12] or carbonyl ligation [13] on the surface of magnetic nanoparticles dramatically affects the magnetic properties. The low temperature magnetization reaches only 75% of the bulk value at low temperature for $NiFe_2O_4$ nanoparticles coated by organic molecules [11]. The chemisorption of CO on the metallic Ni surface leads to the quenching of Ni magnetic moments because electrons of the carbonyl ligation drive the Ni 4$s$ electrons to fill up the 3$d$ shell by repulsive interaction. The Ni atoms in the surface layer, therefore, become nonmagnetic (NM), leaving the magnetism of the inner core unaffected [12, 13]. In many cases, surface magnetic effects are usually featured with a surface spin glass (SG) state and an appreciable reduction of saturation magnetization. For instance, surface SG behaviour has often been observed with ferromagnetic (FM) or ferrimagnetic nanoparticles, such as 6.5 nm $NiFe_2O_4$ [14], 9-10 nm $\gamma$-$Fe_2O_3$ [15], 12 nm Fe nanoparticles [16] and Ni nanochains [17]. Even, the surface SG properties are reported with antiferromagnetic (AFM) $Co_3O_4$ nanowires having a magnetic core-shell structure of $Co_3O_4^{AFM}@Co_3O_4^{SG}$ [18]. Surface magnetism has also been observed with the partially-oxidized composite nanoparticles, $Cu@(Cu_2O^{NM}+CuO^{AFM})$ [19]. Nevertheless, there are other origins leading to surface "dead layer" of magnetic nanoparticles and resulting in the suppression of saturation magnetization [20]. These indicate that the surface magnetic state is one of the most important factors dictating the magnetic properties of nanoparticles.

Ni nanochains with dendritic morphology have been fabricated previously using polyvinyl pyrrolidone (PVP) as a surface modifier [21, 22]. These samples show a SG behavior in addition to a FM phase. The SG behaviors are analyzed from the results of magnetization measurements and are revealed to come from the surface layer [17]. It has been suggested that the surface modifier (PVP) actually tailors the surface magnetic state. In order to further investigate magnetic properties such as the finite size effect of the ferromagnetic (FM) transition point with the Ni nanochains, the diamater-dependent magnetization reversal behavior with the quasi-one dimensional chain-like nanostructure, and the possible transport properties with the nanochain structures, it is important to understand the surface magnetic properties and the thickness of the surface layer. We investigate three powder samples of Ni nanochains with magnetically/structurally core-shell structure, synthesized by a wet chemical method using different surface modifying agents. The estimated outer diameter of the three samples is about 30 nm. They include Ni@Ni$_3$C synthesized using trioctylphosphine oxide (TOPO), and Ni nanochains of two kinds synthesized using PVP and hexadecyltrimethyl ammonium bromide (CTAB), forming the magnetic core-shell structure of Ni@Ni$^{SG}$ and Ni@Ni$^{NM}$, respectively. They are thus labeled as S-TOPO, S-PVP and S-CTAB. For the three samples, the cores are all of FM Ni, however, with a surface shell layer of different magnetic properties, i.e., a NM surface shell of Ni$_3$C for S-TOPO, a magnetically dead layer of Ni (also NM) for S-CTAB, and a SG surface shell for S-PVP.

## 2. Sample preparations and characterizations

The detailed processes of synthesis are reported elsewhere. The process and characterizations for S-TOPO are reported in reference [23], S-PVP is Sample B in reference [22]. The process and the chemical reagents used to prepare S-CTAB is the same as that reported in reference [24], however, with a different reaction temperature at 197 ºC. The chemical reagents used for the preparation of S-CTAB are analytical grade without being further purified. In a typical experiment, 0.5 mmol NiCl$_2$·6H$_2$O and 3 mmol CTAB were dissolved in the solvent of 60 ml glycol. Afterwards, the solution of hydrazine hydrate (50% v/v) by 1 ml was dropped into the mixture. When the solution mixed homogeneously, it was heated to the boiling point (~197 ºC), and kept for 5 hours. The as-obtained sample was washed with ethanol and deionized water.

The crystal structures were characterized by powder X-ray diffraction (XRD) using a Rigaku Dmax 2200 X-ray diffractometer with Cu Kα radiation (λ=0.1542 nm). Transmission electron microscopy (TEM) and high-resolution TEM (HRTEM) investigations were carried out by a JEOL JEM-2100F microscope, equipped with EDS (energy dispersive X-ray spectroscopy). The thermogravimetric (TG) analysis was performed using a Diamond thermogravimetric analyzer (Perkin–Elmer instruments) under a stream of air. The product was heated from 50 °C to 600 °C at a scan rate of 10 °C/min.

Figure 1 shows the XRD patterns for S-TOPO, S-PVP and S-CTAB. It reveals that S-PVP and S-CTAB are of pure Ni phase with a face-centered cubic (fcc) structure. The diffraction peaks correspond to the planes of (111), (200) and (220) of fcc Ni (JCPDS 04-0850). Besides the nickel phase, S-TOPO contains the $Ni_3C$ phase also, which has been characterized in detail and reported in [23]. The peaks marked with open triangles could be assigned to $Ni_3C$ (JCPDS 77-0194), corresponding to the planes of (110), (006), (113), (116), (300), and (119) of $Ni_3C$. It is noteworthy that the (111) peak of Ni overlaps the (113) peak of $Ni_3C$.

Figures 2a, 2b, and 2c show the TEM images of S-TOPO, S-PVP, and S-CTAB, respectively. The morphology is similar, showing chain-like shape with dendritic

structure. Their average diameter is estimated about 30 nm. The corresponding insets show the magnified images. In the inset of figure 2a, a HRTEM image of Ni@Ni$_3$C with a core-shell structure is presented. There is an almost invisible and very thin capped layer outside the Ni$_3$C shell. It is known that a residual nonmagnetic, organic capped layer, TOPO in this case, is inevitable even after several times of thorough rinsing. About 10% mass ratio of the organic capped layer is determined by the TG measurement described below. The Ni$_3$C shell is estimated from the inset about 2 to 4 nm in thickness, which is slightly larger than the average thickness of about 2 nm determined by the magnetic measurements. In the inset of figure 2b for S-PVP, the HRTEM image reveals that the Ni chains are covered with a vague layer of organic remnants. A layer of almost invisible organic remnants is also observed for S-CTAB, as shown in the inset of figure 2c. To further confirm the composition of the samples, EDS measurements were conducted, showing pure Ni element without any other magnetic elements.

The mass ratio of the nonmagnetic organic capped layer is determined by TG measurements, as shown in figure 3. The mass of the as-prepared samples at room temperature is denoted as $m_0$. By heating up the samples gradually, the mass first decrease slightly due to the burning of the organic remnants. Then, the mass increases dramatically arising from the oxidation of Ni, forming NiO, denoted as $m$(NiO). The mass of Ni is then calculated according to the relative ratio of the formula weight, $m$(Ni) = (59/75) $m$(NiO). The correction factor, $m$(Ni)/$m_0$, is then determined as 90%, 87%, and 93% for S-TOPO, S-PVP, and S-CTAB, respectively. The mass correction factor has been accounted for in the normalization of the measured magnetization. It is noted that the mass of C atoms in the $Ni_3C$ shell layer is not taken into account in the above estimation. The NM shell of $Ni_3C$ is treated as if it were a magnetically dead layer of Ni. This introduces about 2% error in mass for $Ni_3C$ being treated as Ni.

3. **Magnetic measurements and analysis**

The magnetization measurements were performed by a Quantum Design SQUID magnetometer. The measurements include, a) temperature dependent saturation magnetization, $M_S(T)$, recorded in the field of 20 kOe from $T = 380$ to 5 K, b) zero-field-cooling (ZFC) and field-cooling (FC) measurements, $M_{ZFC}(T)$ and $M_{FC}(T)$ from $T = 5$ to 380 K, and c) field dependent hysteresis loops, $M(H)$, at a series of fixed temperature from $T = 5$ to 380 K.

Figure 4 shows the temperature dependence of saturation magnetization, $M_S(T)$, of the three samples recorded in the applied field of $H_{app} = 20$ kOe. At $T > 80$ K, the three $M_S(T)$ curves nearly collapse. The values of $M_S(T)$ at $T = 300$ K are determined as 37.7 emu/g (S-TOPO), 38.2 emu/g (S-PVP), and 38.3 emu/g (S-CTAB). They account for about 70% of the corresponding bulk value, ~54.2 emu/g, at 300 K [25]. The reduction of the saturation magnetization is attributed to the magnetically inert property of the surface shell. It is noted that the mass effect of the NM organic capped layer has already been corrected for by the TG measurements. Otherwise, the saturation magnetization per unit mass would have been further reduced by 10 to 15%. For S-TOPO, the reduction in the saturation magnetization is obvious because the $Ni_3C$ shell is NM [26]. The average diameter of the FM cores is then estimated as $D_{core} = 26.6$ nm with the $Ni_3C$ shell thickness of about 1.7 nm. It is consistent with the result of shell thickness, 1-4 nm, observed by the HRTEM investigation [23]. For S-PVP and S-CTAB, although there is no obvious shell structure like the $Ni_3C$ shell of S-TOPO, a FM core of Ni with roughly the same diameter, $D_{core} \sim 26$ nm, is also concluded. At $T < 80$ K, $M_S(T)$ of S-PVP increases dramatically, reaching the bulk value of Ni, $\sim 55$ emu/g. It is attributed to the contribution from the surface SG shell [17]. On the other hand, the shell layer of S-CTAB, although of pure Ni, does not show any magnetism at all, very much like S-TOPO with a NM $Ni_3C$ shell layer. Therefore, S-CTAB exhibits properties of a magnetically core-shell structure, $Ni@Ni^{NM}$, with a magnetically "dead shell" of $Ni^{NM}$. The presence of a magnetically "dead layer" has been reported in the early days with the surface of Ni films [27]. In addition, the magnetic moments of the surface Ni are reported to be quenched, e.g., by the carbonyl ligation on the surface [13].

Figure 5 shows the $M_{ZFC}(T)$ and $M_{FC}(T)$ curves. The ZFC curves were recorded in the applied field, $H_{app} = 90$ Oe, in the warming process after the sample was cooled down to $T = 5$ K under zero applied field. For the FC measurement, the procedure of data collection was the same except that the sample was cooled down to 5 K in the applied magnetic field of 20 kOe. For all of the three samples, the $M_{FC}(T)$ and $M_{ZFC}(T)$ curves separate from each other with the temperature going up to 380 K. It indicates that the blocking temperature is higher than 380 K. The inset shows the blown-up $M_{ZFC}(T)$ curves in the low temperature region. A freezing peak appears at $T_F \sim 8$ K with S-PVP, similar to those at about 13 K observed for 50, 75 and 150 nm Ni nanochains also synthesized using the surfactant of PVP [17]. It further confirms that S-PVP exhibits a magnetically core-shell structure of Ni@Ni[SG]. For the other two samples, S-TOPO and S-CTAB, there is no such characteristic feature, as is expected for the samples with a NM shell layer.

The $M(H)$ measurements for the three samples are performed at various temperature from 5 to 380 K. Figure 6a shows the $M(H)$ curves measured at $T = 5$ K. The magnetization determined in the high field region at $H = 10$ kOe, and the coercivity, $H_C$, are 45.0 emu/g and 600 Oe for S-TOPO, 52.3 emu/g and 568 Oe for S-PVP, and 43.1 emu/g and 634 Oe for S-CTAB, respectively. The magnetization of S-PVP is higher than that of S-TOPO, or S-CTAB by about 20%, attributed to the contribution from the SG shell. The $M(H)$ curves at $T = 300$ K is shown in Figure 6b. The difference in saturation magnetization is more or less reduced. The coercivity determined from the $M(H)$ curves at various temperatures is shown in figure 6c. The magnetization reversal can be described by the fanning mode based on the chain of sphere model proposed by Jacobs and Bean [28], as discussed in our previous work for S-TOPO [23].

## 4. Conclusion

We have investigated the magnetic properties of three samples of Ni nanochains, with an estimated outer diameter of 30 nm, synthesized by different surface modifying agents, including CTAB, TOPO and PVP. The surface magnetic properties are modified significantly and differently by the surfactants in the synthesis process, forming different magnetic shell structures. The CTAB leads to the nanochains of

Ni@Ni$^{NM}$ with a magnetically dead layer of Ni surface shell, while the TOPO makes the final products of Ni@Ni$_3$C nanochains with a shell of NM Ni$_3$C. With PVP, a SG shell layer of Ni is formed, resulting in the magnetic structure of Ni@Ni$^{SG}$. The saturation magnetization of these samples accounts for only 70% of the bulk value at 300 K attributed to the presence of the magnetically inert layer. This value would be further reduced by more than 10% if the mass of the outermost capped layer of organic remnants is not corrected for. The thickness of the shell layer is determined as about 2 nm for all of the three samples.

**Acknowledgements**

This work is supported by the NSFC (Grant No.10874006 and 50725208), State Key Project of Fundamental Research for Nanoscience and Nanotechnology (2006CB932301), National Basic Research Program of China (Nos. 2009CB939901 and 2010CB934601), and Specialized Research Fund for the Doctoral Program of Higher Education (20060006005) as well as by the Innovation Foundation of BUAA for PhD Graduates.

Figure captions

Figure 1. XRD patterns for samples modified with TOPO (S-TOPO), PVP (S-PVP) and CTAB (S-CTAB).

Figure 2. Morphology revealed by TEM images for (a) S-TOPO, (b) S-PVP and (c) S-CTAB. The insets are the corresponding HRTEM images.

Figure 3. Thermogravimetric analyses for the samples of S-TOPO, S-PVP and S-CTAB. The mass of the as-prepared sample is denoted as $m_0$, and the final NiO is denoted as $m_{NiO}$. The calculated Ni contents, $m_{Ni}$, is 93%, 90% and 87% for TOPO, CTAB and PVP, respectively.

Figure 4. Saturation magnetization, $M_S(T)$, recorded in the applied field of 20 kOe from 380 to 5 K. The magnetization of S-PVP increases dramatically at $T < 80$ K.

Figure 5. ZFC and FC $M(T)$ curves for S-TOPO, S-PVP and S-CTAB measured in the applied field of 90 Oe from 5 K to 380 K. The inset shows the ZFC curves in the low temperature region. The peak around 8 K in the ZFC curve with S-PVP is attributed to the freezing of the surface SG shell.

Figure 6. $M(H)$ curves for S-TOPO, S-PVP and S-CTAB at (a) $T = 5$ K, (b) $T = 300$ K, and (c) coercivity determined from the $M(H)$ curves at different temperatures.

Figures

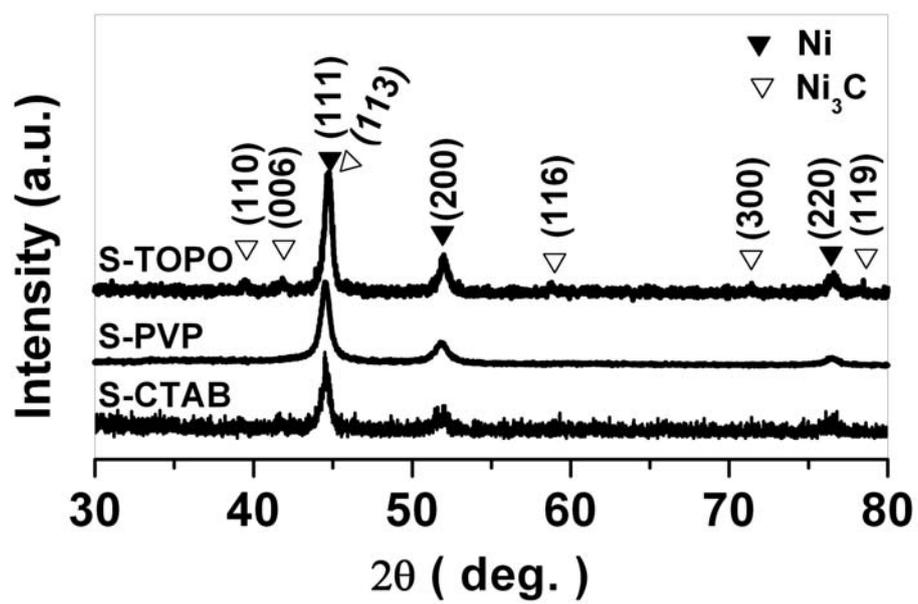

**Figure 1**

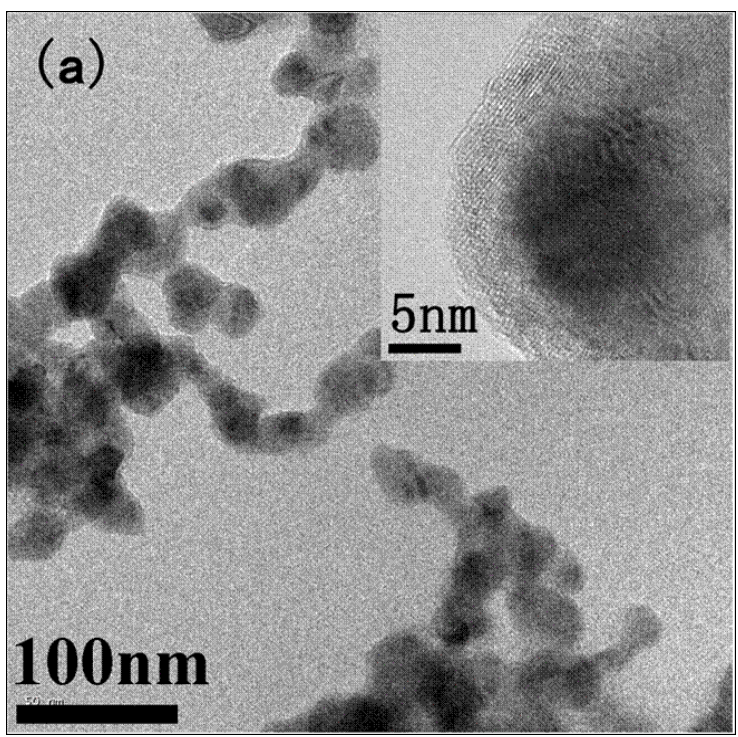

**Figure 2a**

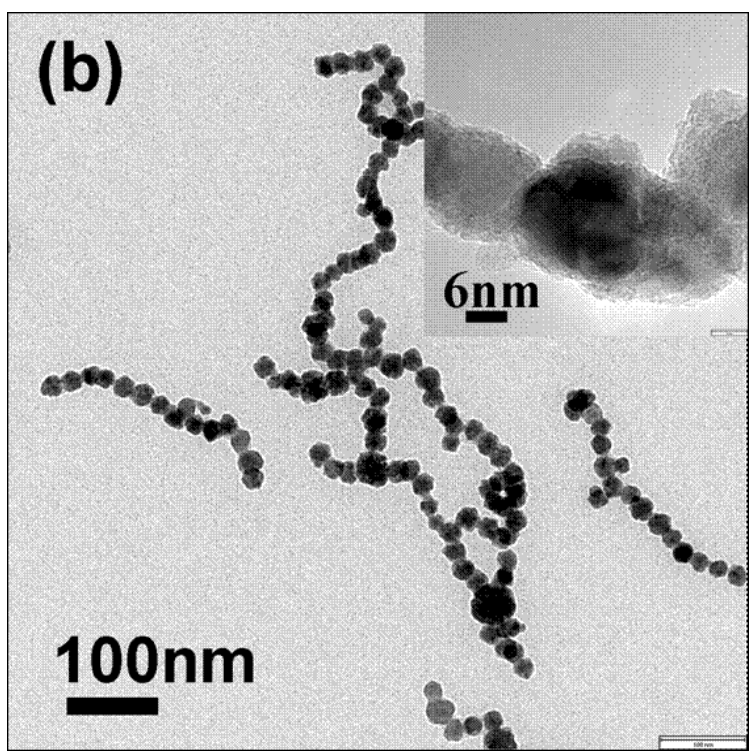

**Figure 2b**

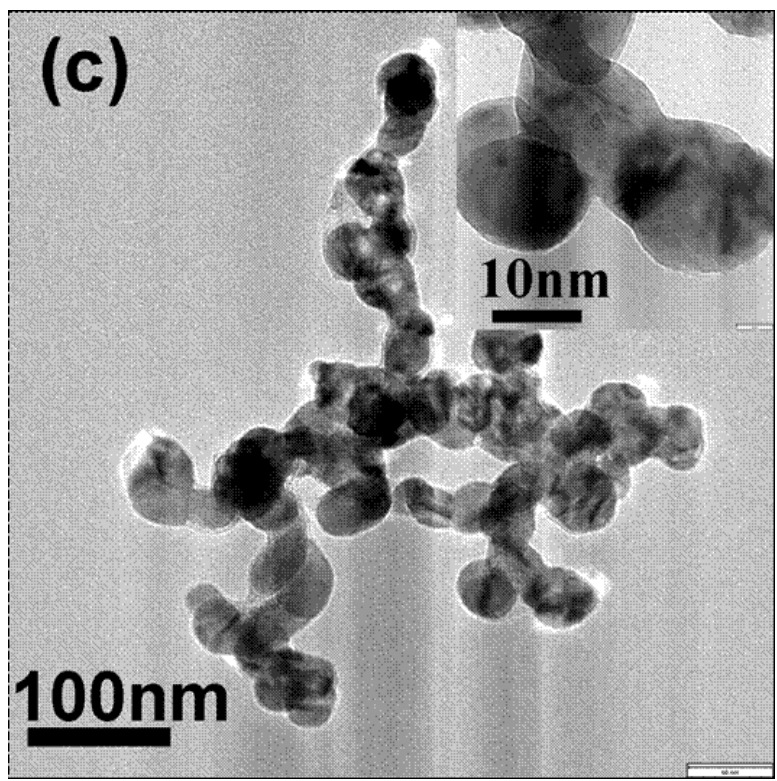

**Figure 2c**

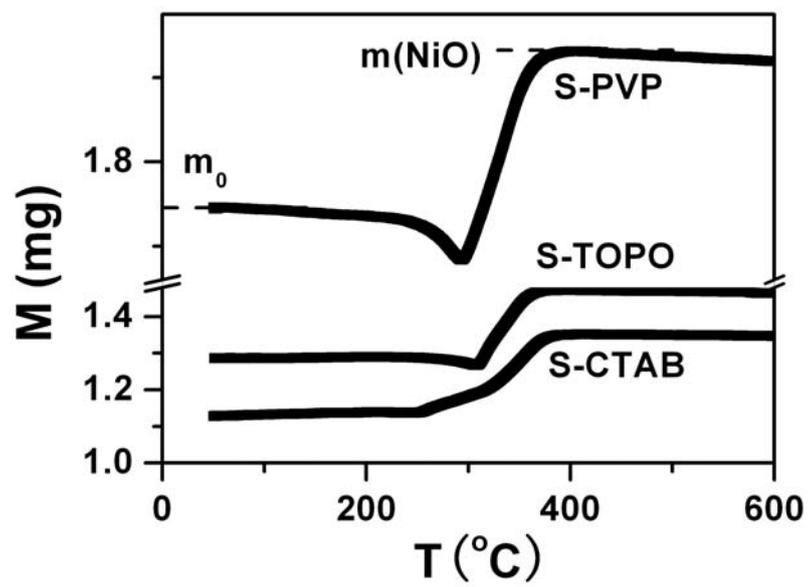

**Figure 3**

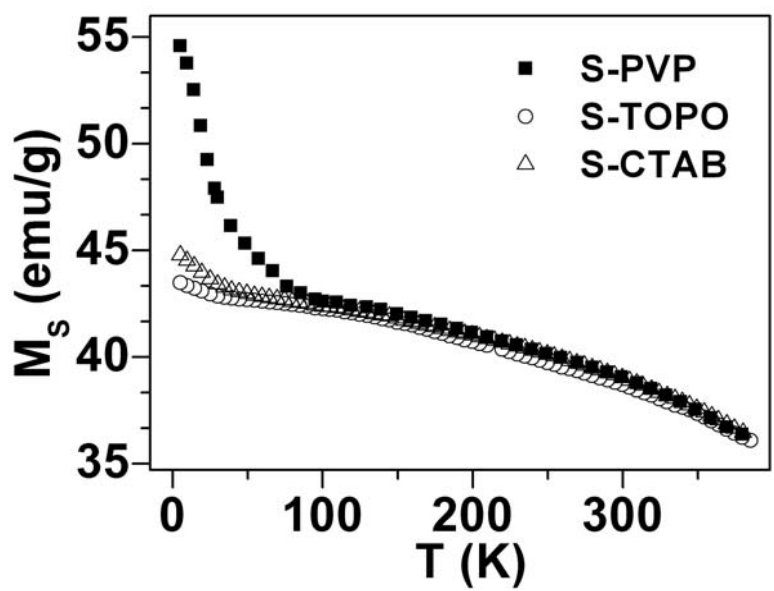

**Figure 4**

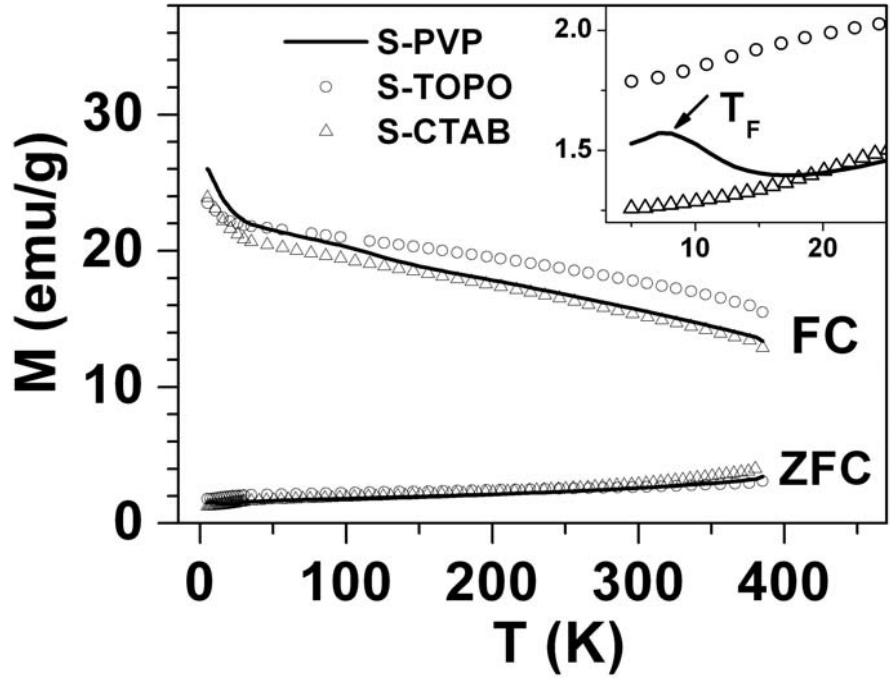

**Figure 5**

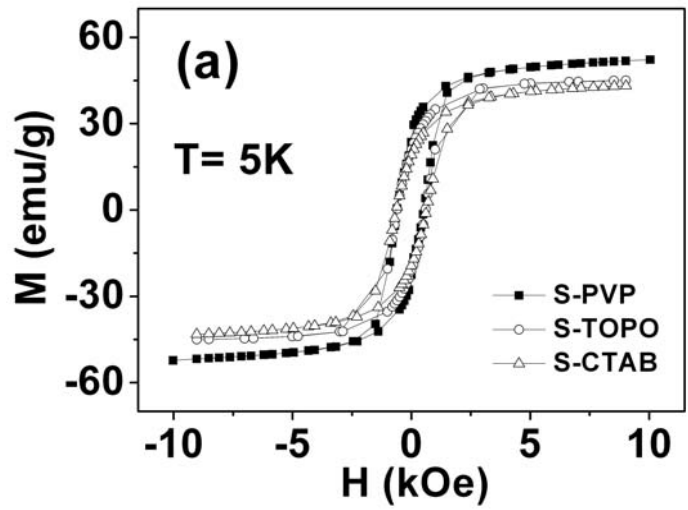

**Figure 6a**

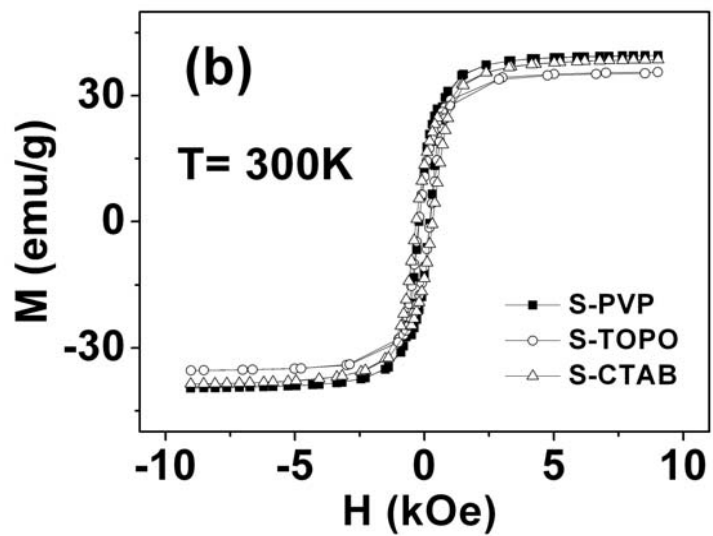

**Figure 6b**

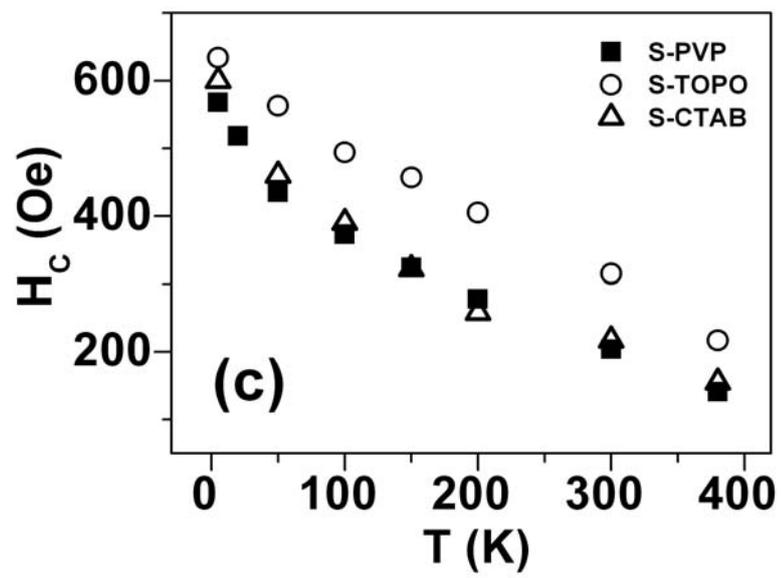

**Figure 6c**